\documentclass[aps,prd,twocolumn,floatfix,nofootinbib,showpacs,superscriptaddress,tightenlines]{revtex4}

\usepackage{amsmath}
\usepackage{lipsum}
\usepackage{amssymb}
\usepackage{amsthm}
\usepackage{bbold}
\usepackage{dcolumn}
\usepackage{epsfig}
\usepackage{graphics}
\usepackage{graphicx}
\usepackage{longtable}
\usepackage{color}

\usepackage{xspace}
\usepackage{cancel}

\newcommand{\me}{\mathrm{e}} 
\newcommand{\mi}{\mathrm{i}} 

\newcommand{\GeV}{\text{GeV}} 

\newcommand{\ie}{\text{i.e.}\xspace}

\newcommand{\eqn}[1]{Eq.~(\ref{#1})}
\newcommand{\fig}[1]{Fig.~\ref{#1}}
\newcommand{\tab}[1]{Table~\ref{#1}}
\newcommand{\sect}[1]{Section~\ref{#1}}

\newcommand{\figs}[2]{Figs.~(\ref{#1},\ref{#2})}

\newcommand{\eqnss}[2]{Eqs.~(\ref{#1},\ref{#2})}

\newcommand{\tabss}[2]{Tables~(\ref{#1},\ref{#2})}

\newcommand{\df}[1]{\hspace{-0.2em}\ensuremath{\frac{\mathrm{d}^{4}#1}{(2\pi)^{4}}}\,}
\newcommand{\dx}[1]{\hspace{-0.2em}\ensuremath{\mathrm{d}#1}\,}

\newcommand{\Tr}{\text{Tr}}

\begin{document}

\title{$\eta_{c}$ Elastic and Transition Form Factors: Contact Interaction and Algebraic Model}

\author{Marco A. Bedolla}
\author{Kh\'epani Raya}
\author{J.J. Cobos-Mart\'{\i}nez}
\author{Adnan Bashir}
\affiliation{
Instituto de F\'{i}sica y Matem\'aticas, Universidad Michoacana de San
Nicol\'as Hidalgo, Edificio C-3, Ciudad Universitaria,
Morelia, Michoac\'an 58040, Mexico}

\date{\today}

\begin{abstract}

For the flavor-singlet heavy quark system of charmonia in the
pseudoscalar ($\eta_c(1S)$) channel, we calculate the elastic
(EFF) and transition form factors (TFF) ($\eta_c(1S) \rightarrow
\gamma \gamma^*$) for a wide range of photon momentum transfer
squared ($Q^2$). The framework for this analysis is provided by a
symmetry-preserving Schwinger-Dyson equation (SDE) and
Bethe-Salpeter equation (BSE) treatment of a vector$\times$vector
contact interaction (CI). We also employ an algebraic model (AM),
developed earlier to describe the light quark systems. It
correctly correlates infrared and ultraviolet dynamics of quantum
chromodynamics (QCD).
 The CI results agree with the lattice data for low
$Q^2$. For $Q^2 \geqslant Q_0^2$, the results start deviating from
the lattice results by more than $20 \%$. $Q_0^2 \thickapprox 2.5
{\rm GeV}^2$ for the EFF and $\thickapprox 25 {\rm GeV}^2$ for the
TFF. We also present the results for the EFF, TFF as well as
$\eta_c(1S)$ parton distribution amplitude for the AM. Wherever
the comparison is possible, these results are in excellent
agreement with the lattice, perturbative QCD, the results obtained
through an SDE-BSE study, employing refined truncations, as well
as the experimental findings of the BABAR experiment.

\end{abstract}


\maketitle

\date{\today}

\section{\label{sec:intro}Introduction}


The internal dynamics of mesons, orchestrated by quantum
chromodynamics (QCD), dictates their observable properties.
Electromagnetic elastic (EFF) and transition form factors (TFF)
provide important examples. In experiment, these quantities are
extracted through the meson interaction with a virtual photon
which probes them at different resolution scales. Several
experimental set ups such as the BABAR, Belle and the upcoming
Belle II and the 12 GeV upgrade of the Jefferson Laboratory hold
the potential to measure these form factors for a large range of
probing photon virtualities. For example, for the $\eta_c(1S)
\rightarrow \gamma \gamma^*$ transition form factor, BABAR has
provided us with results in the range of $ 0 \lessapprox Q^2
\lessapprox 40{\rm GeV}^2$.

Within the well-established framework of Schwinger-Dyson (SDE) and
Bethe-Salpeter equations (BSE), we can investigate the
nonperturbative dynamics of the bound states through first
principles in the continuum. SDEs for QCD have been extensively
applied to deepen our understanding of the light
quark~\cite{Jain:1993qh,Maris:1997hd,Maris:1999nt} and gluon
propagators~\cite{Boucaud:2008ky,Aguilar:2008xm,Pennington:2011xs},
quark-gluon and quark-photon
interactions~\cite{Chang:2009zb,Kizilersu:2009kg,
Bashir:2011dp,Gomez-Rocha:2016cji,Aslam:2015nia,Bashir:2011ij,Bashir:2011vg,Bashir:2004hh},
meson spectra below the masses of 1 GeV as well as their static
and dynamic properties. The form factors for the light mesons
through such studies have been reported in the
Refs.~\cite{Maris:2000sk,Maris:2002mz,Bhagwat:2006pu,Huber:2008id,GutierrezGuerrero:2010md,Roberts:2010rn,Roberts:2011wy,Raya:2015gva}.

The problem of heavy meson's static properties has been addressed
within a consistent rainbow-ladder (RL) truncation of the SDE-BSE
kernels with varying degree of sophistication for the interaction
kernels in
~\cite{Jain:1993qh,Krassnigg:2004if,Bhagwat:2006xi,Maris:2005tt,Blank:2011ha,
Maris:2006ea,Krassnigg:2009zh,Bhagwat:2004hn,Souchlas:2010zz,Rojas:2014aka,
Bedolla:2015mpa,Fischer:2014cfa,Kubrak:2014ela,Hilger:2014nma} as
well as in the lattice-regularized QCD~\cite{Dudek:2006ej}.
Furthermore, within the lattice QCD approach, radiative
transitions and two photon decays of charmonium have been computed
recently with a satisfactorily agreeable comparison with
experimental data~\cite{Dudek:2007zz,Dudek:2006ej,Dudek:2006ud,
Druzhinin:2010zza}.


The extension of the above program to the form factors of heavy
mesons in the SDE-BSE approach is not straightforward. It becomes
numerically cumbersome as the quark propagator has to be sampled
in a large region of the complex plane. However, a few years ago,
a simple alternative model was crafted to have a qualitative
guideline to study light meson properties. It was assumed that the
quarks interact, not via massless vector-boson exchange, but
instead through a symmetry preserving vector-vector contact
interaction (CI)~\cite{GutierrezGuerrero:2010md,Roberts:2010rn,
Chen:2012qr,Roberts:2011cf,Roberts:2011wy}. This interaction is
capable of providing a good description of the meson and baryon
ground and excited-states masses for light
quarks~\cite{GutierrezGuerrero:2010md,
Roberts:2010rn,Chen:2012qr,Roberts:2011cf}. The results obtained
for the static properties through the CI are also quantitatively
comparable to those arrived at by employing sophisticated model
interactions which mimic QCD
closely~\cite{Bashir:2012fs,Eichmann:2008ae,Maris:2006ea,
Cloet:2007pi}. The form factors are expectedly harder but a
qualitative guide is important to make comparison and contrast
with real QCD predictions and experiment.

In a previous work~\cite{Bedolla:2015mpa}, we extended this CI
model to the heavy quark sector to obtain the mass spectrum of
charmonia and the decay constants for the pseudoscalar and vector
channels. In most cases, the agreement achieved with available
experimental data was gratifying. The present article applies this
model, with exactly the same input parameters, to the computation
of form factors associated with the processes
$\eta_{c}\gamma^{*}\to\eta_{c}$ and $\eta_{c}\to\gamma\gamma^{*}$,
namely $\eta_{c}$ EFF and its TFF to $\gamma\gamma^{*}$,
respectively.

In addition to the CI, we also employ an SDE-based algebraic model
(AM), introduced in Ref.~\cite{Chang:2013pq} and refined later in
Ref.~\cite{Raya:2015gva}. This model was constructed in the light
quarks sector to capture both the infrared and ultraviolet
dynamics of QCD in a single, simple and algebraically maneuverable
formalism. It has been successfully employed to gain insight into
the internal nonperturbative dynamics of the charged and neutral
pions~\cite{Chang:2013pq,Chang:2014lva,Mezrag:2014jka,Raya:2015gva}.
We extend it to the case of $\eta_{c}$, calculating the elastic
and transition form factors, as well as the $\eta_{c}$ parton
distribution amplitude (PDA), achieving remarkable agreement with
lattice QCD as well as experiment, whenever possible.

This paper is organized as follows. In \sect{sec:sde-bse}, we
present the necessary SDE-BSE tools and ingredients to study
mesons and compute the EFF and the TFF of charmonia. We summarize
the main features of the CI model for the sake of completeness. We
also introduce the AM and present its extension for charmonia.
\sect{sec:etac-eff} and \sect{etac-tff} have been dedicated to the
computation of $\eta_c$ EFF and TFF with the CI and AM, providing
comparison with lattice QCD, other models and existing
experimental results wherever possible. In~\sect{sec:PDA}, we
calculate the $\eta_{c}$ PDA through the AM and compare it with
the lattice QCD and perturbative calculations. Finally, in
\sect{sec:conclusions}, we present our conclusions.

\section{\label{sec:sde-bse} The Formalism}

Meson bound states appear as poles in a four-point function. The
condition for the appearance of such a pole in a particular
$J^{PC}$ channel is given by the
BSE~\cite{Gross:1993zj,Salpeter:1951sz, GellMann:1951rw}
 \begin{equation}
 \label{eqn:bse}
 \left[\Gamma_{H}(p;P)\right]_{tu}=
 \int\df{q}K_{tu;rs}(p,q;P)\chi(q;P)_{sr},
 \end{equation}
 where
$\chi(q;P)=S_{f}(q_{+})\Gamma_{H}(q;P)S_{g}(q_{-})$; $q_{+}=q+\eta
P$, $q_{-}=q-(1-\eta)P$; $p$ ($P$) is the relative (total)
momentum of the quark-antiquark system; $S_{f}$ is the $f$-flavor
quark propagator; $\Gamma_{H}(p;P)$ is the meson Bethe-Salpeter
amplitude (BSA), where $H$ specifies the quantum numbers and
flavor content of the meson; $r,s,t,u$ represent color, flavor,
and spinor indices; and $K(p,q;P)$ is the quark-antiquark
scattering kernel. For a comprehensive recent review of the
SDE-BSE formalism and its applications to hadron physics, see for
example Refs.~\cite{Bashir:2012fs,Aznauryan:2012ba}.

The $f$-flavor dressed-quark propagator $S_{f}$ that enters
\eqn{eqn:bse} is obtained as the solution of the quark
SDE~\cite{Roberts:2007jh,Holl:2006ni,Maris:2003vk,Alkofer:2000wg}
 \begin{eqnarray}
 \label{eqn:quark_sde}
 &&\hspace{-0.5cm} S_{f}^{-1}(p)=i\gamma\cdot p + m_{f} + \Sigma_{f}(p), \\
 \label{eqn:quark_se}
 && \hspace{-0.5cm} \Sigma_{f}(p)=\int\df{q}g^{2}D_{\mu\nu}(p-q)\frac{\lambda^{a}}{2}
 \gamma_{\mu}S_{f}(q)\Gamma^{a}_{\nu}(p,q),
\end{eqnarray}
where $g$ is the strong coupling constant, $D_{\mu\nu}$ is the
dressed-gluon propagator, $\Gamma^{a}_{\nu}$ is the
dressed-quark-gluon vertex, and $m_{f}$ is the $f$-flavor
current-quark mass. Since the CI, to be defined later, is
nonrenormalizable, it is not necessary to introduce any
renormalization constant. The chiral limit is obtained by setting
$m_{f}=0$~\cite{Roberts:2007jh,Holl:2006ni,Maris:2003vk}.

Both $D_{\mu\nu}$ and $\Gamma^{a}_{\nu}$ satisfy their own SDE,
which in turn are coupled to the equations containing higher
$n$-point functions and so on {\it ad infinitum}. Therefore, the
quark SDE, \eqn{eqn:quark_sde}, is only one of the infinite set of
coupled nonlinear integral equations. A tractable problem is
defined once we have spelled out our truncation scheme, \ie, once
the gluon propagator and the quark-gluon vertex are specified.

\subsection{\label{sec:rl-ci} The Contact Interaction}

It has been shown in
Refs.~\cite{GutierrezGuerrero:2010md,Roberts:2010rn,
Chen:2012qr,Roberts:2011cf} that a momentum-independent
vector$\times$vector CI is capable of providing a description of
light pseudoscalar and vector mesons static properties,
quantitatively comparable to that obtained using more refined QCD
model interactions~\cite{Bashir:2012fs,Eichmann:2008ae,
Maris:2006ea,Cloet:2007pi}. Furthermore, the $\pi$ and $\rho$ EFF
\cite{GutierrezGuerrero:2010md,Roberts:2011wy} and TFF of the
$\pi$,~\cite{Roberts:2010rn} have also been calculated in this
interaction. In a previous work~\cite{Bedolla:2015mpa}, we have
employed this interaction to evaluate the mass spectrum of
charmonia and the decay constants of the pseudoscalar and vector
meson channels. In this article, we apply it to the computation of
the EFF and TFF of the $\eta_c$. Therefore, we use
 \begin{eqnarray}
 \label{eqn:contact_interaction}
 g^2 D_{\mu \nu}(k) =\frac{4\pi\alpha_{\text{IR}}}{m_g^2}\delta_{\mu \nu} \equiv
 \frac{1}{m_{G}^{2}}\delta_{\mu\nu},
 \end{eqnarray}
 where $m_g=800$ MeV is a gluon mass scale which is in
fact generated dynamically in QCD (see for example
Ref.~\cite{Boucaud:2011ug}) and $\alpha_{\text{IR}}$  is a
parameter that determines the interaction strength. For the
quark-gluon vertex, the rainbow truncation will be used:
 \begin{equation}
 \label{eqn:quark_gluon_vertex_rl}
 \Gamma^{a}_{\mu}(p,q)=\frac{\lambda^{a}}{2}\gamma_{\mu}.
 \end{equation}
 Once the elements of the kernel in the quark SDE have
been specified, we can proceed to generate and analyze its
solution. The general form of the $f$-flavored dressed quark
propagator, the solution of \eqn{eqn:quark_sde}, can be written in
terms of two Lorentz-scalar dressing functions in the following
equivalent and convenient forms:
 \begin{eqnarray}
 \label{eqn:gen_quark_inverse}
 S^{-1}_{f}(p)&=& \mi\gamma\cdot pA_{f}(p^{2}) + B_{f}(p^{2}) \\
 &=& Z^{-1}_{f}(p^{2})\left(i\gamma\cdot p + M_{f}(p^{2})\right).
 \end{eqnarray}
In the latter expression, $Z_{f}(p^{2})$ is known as the
wave-function renormalization, and $M_{f}(p^{2})$ is the dressed,
momentum-dependent quark mass function, which connects current and
constituent quark
masses~\cite{Roberts:2007jh,Holl:2006ni,Maris:2003vk}.

Using \eqnss{eqn:contact_interaction}{eqn:quark_gluon_vertex_rl},
the quark SDE equation can be written as:
 \begin{equation}
 \label{eqn:quark_sde_contact}
 S^{-1}_{f}(p)=i\gamma\cdot p + m_{f}
 + \frac{4}{3}\frac{1}{m_{G}^{2}}\int\df{q}\gamma_{\mu}S_{f}(q)\gamma_{\mu}.
 \end{equation}
The solution of \eqn{eqn:quark_sde_contact} now has the form
 \begin{equation}
 \label{eqn:quark_inverse_contact}
 S_{f}^{-1}(p)= \mi\gamma\cdot p + M_{f}.
 \end{equation}
In other words, for the CI, $Z_{f}(p^{2})=1$ and $M_{f}$ is
momentum independent. It is because the last term on the
right-hand side of \eqn{eqn:quark_sde_contact} is independent of
the external momentum. The mass $M_{f}$ is determined as the
solution of
 \begin{equation}
 \label{eqn:const_mass} M_{f} = m_{f} +
 \frac{16M_{f}}{3\pi^{2}m_{G}^{2}}\int\df{q}\frac{1}{q^{2}+M_{f}^{2}}.
 \end{equation}
Since \eqn{eqn:const_mass} is divergent, we must adopt a
regularization procedure. We employ the proper time regularization
scheme~\cite{Ebert:1996vx} and write
 \begin{eqnarray}
 \label{eqn:regularisation_procedure}
 \frac{1}{q^{2}+M^{2}}&=&\int_{0}^{\infty}\dx{\tau}\me^{-\tau(q^{2}+M^{2})}
 \to \int_{\tau_{\text{UV}}^{2}}^{\tau_{{\text{IR}}^{2}}}\dx{\tau}\me^{-\tau(q^{2}+M^{2})}
 \nonumber \\
 &=&\frac{\me^{-\tau_{\text{UV}}^2(q^{2}+M^{2})}-\me^{-\tau_{\text{IR}}^2(q^{2}+M^{2})}}{q^{2}+M^2},
\end{eqnarray}
 where $\tau_{\text{IR}}^{2}$ and $\tau_{\text{UV}}^{2}$ are, respectively,
infrared and ultraviolet regulators. Note that a nonzero value for
$\tau_{\text{IR}}\equiv 1/\Lambda_{\text{IR}}$ implements
confinement by ensuring the absence of quark production
thresholds~\cite{Roberts:2007ji}. Furthermore, since
\eqn{eqn:contact_interaction} does not define a renormalizable
theory, $\tau_{\text{UV}}\equiv 1/\Lambda_{\text{UV}}$ cannot be
removed. Instead, it plays a dynamical role and sets the scale for
all dimensioned quantities. The importance of an ultraviolet
cut-off in Nambu--Jona-Lasinio type models has also been discussed
in Refs.~\cite{Farias:2005cr,Farias:2006cs}. Thus
 \begin{equation}
 \label{eqn:const_mass_reg} M_{f}= m_{f} +
 \frac{16M_{f}}{3\pi^{2}m_{G}^{2}}\mathcal{I}_{01}(M_{f}^{2};\tau_{\text{IR}},
 \tau_{\text{UV}}),
 \end{equation}
 where
 \begin{equation}
 \label{eqn:Ifun}
 \mathcal{I}_{0n}(M^{2};\tau_{\text{IR}},\tau_{\text{UV}})=
 \frac{\left(M^{2}\right)^{2-n}}{16\pi^{2}\Gamma(n)}
 \Gamma(n-2,\tau_{\text{UV}}^{2}M^{2},\tau_{\text{IR}}^{2}M^{2}),
 \end{equation}
 and $\Gamma(a,z_{1},z_{2})$ is the generalized
incomplete gamma function.

\subsection{\label{sec:axwti} The Axial-Vector Ward-Takahashi Identity}

The phenomenological features of dynamical chiral symmetry
breaking (DCSB) in QCD can be understood  by means of the
axial-vector Ward-Takahashi identity (axWTI). In the chiral limit,
it reads
 \begin{equation}
 \label{eqn:axwti} -\mi P_{\mu}\Gamma_{5\mu}(p;P) =
 S^{-1}(p_{+})\gamma_{5} + \gamma_{5}S^{-1}(p_{-}).
 \end{equation}
The axWTI implies a relationship between the kernel in the BSE,
\eqn{eqn:bse}, and that in the quark SDE, \eqn{eqn:quark_sde},
 \begin{multline}
 \label{eqn:axialwti_kernels_rels}
 \int\df{q}K_{tu;rs}(p,q;P)\left[\gamma_{5}S(q_{-})+ S(q_{+})\gamma_{5}\right]_{sr} \\
 = \left[\Sigma(p_{+})\gamma_{5} + \gamma_{5}\Sigma(p_{-})
 \right]_{tu}.
\end{multline}
 This relation must be preserved by any viable truncation
scheme of the SDE-BSE coupled system, thus constraining the
content of the quark-antiquark scattering kernel $K(p,q;P)$ if an
essential symmetry of the strong interactions, and its breaking
pattern, are to be faithfully reproduced. Satisfying this identity
is particularly important when dynamical chiral symmetry breaking
dominates the physics.

However, from a practical point of view,
\eqn{eqn:axialwti_kernels_rels} provides a way of obtaining the
quark-antiquark scattering kernel, given an expression for the
quark self-energy $\Sigma$. For the CI under study,
\eqn{eqn:axialwti_kernels_rels} can be easily satisfied. The
resulting expression for the quark-antiquark scattering kernel is
the RL truncation. This kernel is the leading-order term in a
nonperturbative, symmetry-preserving truncation scheme, which is
known and understood to  be accurate for the pseudoscalar and
vector mesons. Moreover, it guarantees electromagnetic current
conservation~\cite{Roberts:2007ji}:
 \begin{equation}
 \label{eqn:bskernel_rl_contact} K(p,q;P)_{tu;rs}= -g^{2}D_{\mu\nu}(p-q)
 \left[\frac{\lambda^{a}}{2}\gamma_{\mu}\right]_{ts}
 \left[\frac{\lambda^{a}}{2}\gamma_{\nu}\right]_{ru},
 \end{equation}
where $g^{2}D_{\mu\nu}$ is given by \eqn{eqn:contact_interaction}.
Using the interaction that we have specified via
\eqnss{eqn:contact_interaction}{eqn:quark_gluon_vertex_rl}, the
homogeneous BSE for a meson ($\eta=1$) takes a simple form:
 \begin{equation}
 \label{eqn:bse_contact} \Gamma_{H}(p;P)=-\frac{4}{3}\frac{1}{m_{G}^{2}}\int\df{q}
 \gamma_{\mu}S_{f}(q+P)\Gamma_{H}(q;P)S_{g}(q)\gamma_{\mu}.
 \end{equation}
Since the interaction kernel given in
\eqn{eqn:bskernel_rl_contact} does not depend on the external
relative momentum for the CI, a symmetry-preserving regularization
will yield solutions which are independent of it. With a
dependence on the relative momentum not supported by the CI, the
general form of the BSA for the pseudoscalar and vector channels
is given in Ref.~\cite{LlewellynSmith:1969az}
 \begin{eqnarray}
 \label{eqn:psbsagral}
 \Gamma^{\eta_{c}}(P)&=& \gamma_{5}\left[\mi E^{\eta_{c}}(P)
 + \frac{1}{2M}\gamma\cdot P F^{\eta_{c}}(P)\right], \\
 \label{eqn:vbsagral}
 \Gamma^{J/\Psi}_{\mu}(P)&=&\gamma^{T}_{\mu}E^{J/\Psi}(P)
 + \frac{1}{2M}\sigma_{\mu\nu}P_{\nu}F^{J/\Psi}(P),
 \end{eqnarray}
where $M=M_{c}/2$ is a mass scale, with $M_{c}$ being the solution
of \eqn{eqn:const_mass_reg}. Results for the physical observables
are clearly independent of this choice.

Since the BSE is a homogeneous equation, the BSA has to be
normalized by a separate condition. In the RL truncation of the
BSE, this condition is
 \begin{equation}
 \label{eqn:RLNorm}
 P_{\mu}=N_{c}\frac{\partial}{\partial P_{\mu}}\int\df{q}
 \Tr\left[\overline{\Gamma}_{H}(-Q)S(q_{+})\Gamma_{H}(Q)S(q)\right],
 \end{equation}
at $Q=P$, with $P^{2}=-m_{H}^{2}$ (we choose $\eta=1$) . Equation
(\ref{eqn:RLNorm}) ensures that the residue of the four-point
function at the mass pole is unity. Here, $\Gamma_{H}$ is the
normalized BSA and $\overline{\Gamma}_{H}$ its charge-conjugated
version. For every channel, we will rescale $\Gamma_{H}$ such that
\eqn{eqn:RLNorm} is satisfied. Furthermore, for the vector channel
there is an additional factor of 1/3 on the right hand side to
account for all three meson polarizations.

Once the BSA has been normalized canonically with
\eqn{eqn:RLNorm}, we can calculate observables from it. For
example, the pseudoscalar leptonic decay constant $f_{0^{-}}$ is
defined by
 \begin{equation}
 \label{eqn:psdecaydef}
 P_{\mu}f_{0^{-}}=N_{c}\int\df{q}\Tr\left[\gamma_{5}\gamma_{\mu}
 S(q_{+})\Gamma_{0^{-}}(P)S(q_{-})\right].
 \end{equation}
 Similarly, the vector decay constant $f_{1^{-}}$ is:
 \begin{equation}
 \label{eqn:vdecaydef}
 m_{1^{-}}f_{1^{-}}=\frac{N_{c}}{3}\int\df{q}\Tr\left[\gamma_{\mu}
 S(q_{+})\Gamma^{1^{-}}_{\mu}S(q_{-})\right],
 \end{equation}
where $m_{1^{-}}$ is the mass of the vector bound state, and the
factor of 3 in the denominator comes from summing over the three
polarizations of the spin-1 meson.

\subsubsection{\label{sec:axwti-cor} A Corollary of the Axial-Vector WTI}

There are further non trivial consequences of the axWTI and the
CI. They define our regularization procedure, which must maintain

\begin{eqnarray}
\label{eqn:wticorollary} 0&=&\int\df{q} \left[\frac{P\cdot
q_{+}}{q_{+}^{2}+M_{f}^{2}} - \frac{P\cdot
q_{-}}{q_{-}^{2}+M_{g}^{2}}
\right] \nonumber \\
&=&\int_{0}^{1}\dx{x}\int\df{q} \frac{\frac{1}{2}q^{2} +
\mathfrak{M}^{2}} {\left(q^{2}+ \mathfrak{M}^{2}\right)^{2}},
\end{eqnarray}
 where $\mathfrak{M}^{2}=M_{f}^{2}x + M_{g}^{2}(1-x)+ x(1-x)P^{2}$.
This ensures that~\eqn{eqn:axwti} is satisfied. Equation
(\ref{eqn:wticorollary}) states that the axWTI is satisfied if,
and only if, the model is regularized so as to ensure there are no
quadratic or logarithmic divergences. Unsurprisingly, these are
the circumstances under which a shift in integration variables is
permitted, an operation required in order to prove
\eqn{eqn:axwti}~\cite{GutierrezGuerrero:2010md,
Roberts:2010rn,Chen:2012qr,Roberts:2011cf}. The constraint given
by \eqn{eqn:wticorollary} will be implemented in all our
calculations so that~\eqn{eqn:axwti} is unequivocally preserved.

\subsection{\label{sec:charmonia-ci} The Contact Interaction for
Charmonia}


\begin{table}[h]
\begin{center}
\begin{tabular}{lllll}
\hline \hline
    & \multicolumn{4}{c}{masses}  \\
\hline
& $m_{\eta_{c}(1S)}$ & $m_{J/\Psi(1S)}$ & $m_{\chi_{c_{0}}(1P)}$ & $m_{\chi_{c_{1}}(1P)}$
\\
\hline
Experiment ~\cite{0954-3899-37-7A-075021} & 2.983 & 3.096 & 3.414 & 3.510  \\
Contact Interaction & 2.950$^{*}$ & 3.129 & 3.407 & 3.433 \\
JM ~\cite{Jain:1993qh} & 2.821 & 3.1 & 3.605 & - \\
BK ~\cite{Blank:2011ha} & 2.928 & 3.111 & 3.321 & 3.437 \\
RB1 ~\cite{Rojas:2014aka} & 3.065 & - & - & - \\
RB2 ~\cite{Rojas:2014aka} & 3.210 & - & - & - \\
 FKW ~\cite{Fischer:2014cfa} &  2.925
&  3.113 &  3.323 &  3.489 \\
\hline \hline
\end{tabular}
\caption{\label{tab:mcc_all_opt} Ground state charmonia masses
obtained with the best-fit parameter set: $m_{g}=0.8\,\GeV$,
$\alpha_{IR}=0.93\pi/20$, $\Lambda_{\text{IR}}=0.24\,\GeV$,
$\Lambda_{\text{UV}}=2.788\,\GeV$. The current-quark mass is
$m_{c}=0.956^{*}\,\GeV$, and the dynamically generated
constituent-like mass is $M_{c}=1.497\,\GeV$. Dimensioned
quantities are in GeV. ($^{*}=$ This parameter set was obtained
from the best-fit to the mass and decay constant of the
pseudoscalar and vector channels). The average percentage error,
with respect to experimental data, is $1.14\%$.}
\end{center}
\end{table}

\begin{table}[h]
\begin{center}
\begin{tabular}{lllll}
\hline \hline
    & \multicolumn{3}{c}{decay constants}  \\
\hline \hline
 & $f_{\eta_{c}}$ & $f_{J/\Psi}$
\\
Lattice QCD  & 0.395~\cite{Davies:2010ip} & 0.405~\cite{Donald:2012ga} \\
S1rp ~\cite{Souchlas:2010zz} & 0.239 & 0.198 \\
S3ccp ~\cite{Souchlas:2010zz} & 0.326 & 0.330 \\
BK ~\cite{Blank:2011ha} & 0.399 & 0.448 \\
Contact Interaction & 0.305 & 0.220  \\
\hline \hline
\end{tabular}
\caption{\label{tab:DecayCNew} The decay constants for the states
$\eta_{c}(1S)$ and $J/\Psi(1S)$ obtained with $m_{g}=0.8\,\GeV$,
$\alpha_{IR}=0.93\pi/20$, $\Lambda_{\text{IR}}=0.24\,\GeV$,
$\Lambda_{\text{UV}}=2.788\,\GeV$.
The current-quark mass is $m_{c}=0.956\,\GeV$. Dimensioned quantities are in
GeV.}
\end{center}
\end{table}

In a recent work~\cite{Bedolla:2015mpa}, we have developed a CI
model for charmonia. The results for the low lying mass spectrum
of corresponding mesons are presented in \tab{tab:mcc_all_opt}.
They are in excellent agreement with experimental data (with an
average percentage error of $1.14\%$) and, consequently,  with the
findings of more sophisticated SDE-BSE model
calculations~\cite{Blank:2011ha,Rojas:2014aka,Fischer:2014cfa,
Kubrak:2014ela,Hilger:2014nma} and lattice QCD
computations,~\cite{Follana:2006rc,Wagner:2013laa}. The fact that
a RL truncation with a CI describes the mass spectrum of ground
state charmonia so well can be understood in a simple way: since
the wave function renormalization and quark mass function are
momentum-independent, the heavy-quark--gluon vertex can reasonably
be approximated by a bare vertex. The decay constants calculated
in~\cite{Bedolla:2015mpa} for the $\eta_{c}$ and $J/\Psi$ channels
are given in \tab{tab:DecayCNew}. For the pseudoscalar meson, the
result is in decent agreement with the lattice QCD result. Though
it is not exactly the case for the vector channel, it is one of
the best results in such models (see Ref.~\cite{Bedolla:2015mpa}
for an extended discussion). Note that the results presented
in~\tabss{tab:mcc_all_opt}{tab:DecayCNew} correspond to a minimal
extension of the CI model developed primarily for the light quarks
in~\cite{GutierrezGuerrero:2010md,Roberts:2010rn,Chen:2012qr,
Roberts:2011cf}. A naive application of this earlier model to
quarkonia yielded unacceptable results. The reason can be traced
back to the fact that the decay constant is influenced by the high
momentum tails of the dressed-quark propagator and the BSAs
~\cite{Bhagwat:2004hn,Maris:1997tm,Maris:1999nt}. These tails
probe the wave-function of quarkonia at the origin. Contrastingly,
the CI yields a constant mass with no perturbative tail for large
momenta. Therefore, this artifact of quarkonia had to be built
into the model in an alternative manner. Furthermore, we know
that, as the masses become higher, mesons become increasingly
point-like in configuration space. The closer the quarks get, the
weaker is the coupling strength between them. We thus extended the
CI model by reducing the effective coupling $\alpha_{IR}$,
accompanied by an appropriate increase in the ultraviolet cutoff.
However, we retained the parameters $m_g$ and
$\Lambda_{\text{IR}}$ of the light sector since modern studies of
the gluon propagator indicate that in the infrared, the
dynamically generated gluon mass scale virtually remains
unaffected by the introduction of heavy dynamical quark masses,
see for example Refs.~\cite{Ayala:2012pb,Bashir:2013zha}. In the
subsequent sections, we shall use this extended CI to evaluate the
$\eta_c$ EFF and TFF to $\gamma^* \gamma$.

\subsection{\label{sec:charmonia-am} The Algebraic Model for
Charmonia}

We also consider the following SDE-based algebraic model, with a
simple extension to the heavy quark sector:
\begin{eqnarray}
\label{eqn:algmod}
S^{-1}(p)&=& \mi\gamma\cdot p + M \nonumber, \\
\rho_{\nu}(z)&=&\frac{\Gamma(\frac{3}{2}+\nu)}{\Gamma(\frac{1}{2})\Gamma(1+\nu)}
(1-z^{2})^{\nu}, \nonumber \\
\Gamma_{\eta_{c}}(k;P)&=&\mi\gamma_{5}\mathcal{N}\frac{M}{f_{\eta_{c}}}
\int_{-1}^{1}\dx{z}\rho_{\nu}(z)\frac{M^{2}}{(k+z\sigma
P/2)^{2}+M^{2}}, \nonumber \\ \label{Charmonia-am}
\end{eqnarray}
 where $P^{2}=-m_{\eta_{c}}^{2}$. $\mathcal{N}$ plays the
role of the canonical normalization condition,~\eqn{eqn:RLNorm},
$M$ is fixed such that $f_{\eta_{c}}=0.361\,\GeV$ and
$\sigma=m_{\pi}/m_{\eta_{c}}$. A small $\sigma$ suppresses the
angular dependence $k\cdot P$ of the BSA, characteristic of heavy
mesons, while $\sigma=1$ recovers the SDE-based AM for the
pion~\cite{Raya:2015gva}.

The parameter $\nu$ that appears in \eqn{Charmonia-am} strongly
influences the form of the resulting parton distribution amplitude
(PDA)~\cite{Raya:2015gva}. For the pion, $\nu=1$ produces
$\phi_{\pi}\sim x(1-x)$. This expression is in agreement with the
asymptotic QCD prediction. $\nu=-1/2$ yields $\phi_{\pi}\sim
\sqrt{x(1-x)}$,
 in keeping with a realistic PDA at the hadronic
scale~\cite{Chang:2013pq,Cloet:2013tta, Braun:2006dg}. As the PDA
of the pion plays a crucial role in determining the asymptotic
behavior of its EFF and the TFF to $\gamma^* \gamma$, the AM is an
efficient model to encode both its nonperturbative and asymptotic
dynamics, as exemplified in Ref.~\cite{Raya:2015gva}. This model
has also been used to calculate pion's valence dressed-quark
generalized parton distribution (GPD) $H^v_{\pi}(x,\xi,t)$ for
``skewness" $\xi=0$,~\cite{Mezrag:2014jka}.

Owing to the above discussion about the form of the pion PDA and
its relation to the $\nu$ parameter, we fix $\nu=1$ for the AM to
study the $\eta_c$,~\eqn{eqn:algmod}. Once the AM parameters have
been fixed, through the values of $m_{\eta_{c}}, f_{\eta_{c}}$,
and $\nu$, one can use it to calculate the EFF and TFF for
$\eta_c$.

\subsection{\label{sec:etac-numres} Electromagnetic Interaction:
The Quark-Photon Vertex}

The interaction of a virtual photon with a meson probes its
internal structure and dynamics. The impulse approximation allows
electromagnetic processes to be described in terms of quark
propagators, bound state BSAs, and the quark-photon vertex. In
combination with the RL truncation for the quark propagator and
vertices, it ensures electromagnetic current
conservation~\cite{Hawes:1998bz,Maris:1999bh,Maris:2000sk,Maris:2002na,
Bhagwat:2006pu}. Phenomenologically, this approximation has proved
to be very successful in describing EFF and TFF of light
pseudoscalar and vector mesons~\cite{Maris:2000sk,Bhagwat:2006pu,
Maris:2002mz}.

The coupling of a photon with the bound state's charged
constituent is given by the quark-photon vertex. In addition to
being determined by its own SDE, which is highly nontrivial to
solve, the quark-photon vertex $\Gamma_{\mu}(p_{+},p_{-};Q)$ is
constrained by the gauge invariance of quantum electrodynamics
(QED) through the vector Ward-Takahashi identity (WTI)
\begin{equation}
\label{eqn:vwti}
\mi Q_{\mu}\Gamma_{\mu}(p_{+},p_{-};Q)=S^{-1}(p_{+}) - S^{-1}(p_{-}).
\end{equation}
\noindent Preserving this identity, and its $Q\to 0$ limit, is key
to the conservation of electromagnetic current. In our present
truncation, the SDE for the quark-photon vertex, consistent with
\eqn{eqn:bskernel_rl_contact} (truncated at the RL level), is
\begin{equation}
\label{eqn:quark_photon_sde_rl}
\Gamma_{\mu}(p;Q)=\gamma_{\mu}-\frac{4}{3}\frac{1}{m_{G}^{2}}
\int\df{q}\gamma_{\alpha}S(q_{+})\Gamma_{\mu}(q;Q)S(q_{-})\gamma_{\alpha},
\end{equation}
\noindent where $q_{+}=q+Q$, and $q_{-}=q$. Noting that the
right-hand-side is independent of the relative momenta, the
general form of the quark-photon vertex is
\begin{equation}
\label{eqn:quark_photon_rl}
\Gamma_{\mu}(Q)=\gamma_{\mu}^{T}P_{T}(Q^{2}) + \gamma_{\mu}^{L}P_{L}(Q^{2}),
\end{equation}
\noindent where $Q_{\mu}\gamma_{\mu}^{T}=0$,
$\gamma_{\mu}^{T}+\gamma_{\mu}^{L}=\gamma_{\mu}$. Furthermore,
note that with the usage of \eqn{eqn:quark_inverse_contact} and
\eqn{eqn:quark_photon_rl}, the vector WTI is trivially obeyed.
Moreover, the bare vertex $\gamma_{\mu}$ also satisfies the WTI
for the contact interaction propagator
\eqn{eqn:quark_inverse_contact}. However, the bare vertex does not
contain vector meson poles, which are relevant for the correct
description of the charge interaction radius; see for
example~\cite{Maris:1999bh}.

Taking appropriate Dirac traces, and using the constraint of
\eqn{eqn:wticorollary}, which stems from the axWTI, we find
$P_{L}(Q^{2})=1$ and
\begin{equation}
\label{eqn:FTFinal}
P_{T}(Q^{2})=\frac{1}{1-K_{J/\Psi}(Q^{2})},
\end{equation}
\noindent where $K_{J/\Psi}$ is the Bethe-Salpeter bound state
kernel in the vector channel, \eqn{eqn:vbsagral}, in the present
truncation~\cite{Bedolla:2015mpa}. Thus, because of the dressing
of the quark-photon vertex, our form factors (EFF and TFF) will
have a pole at $Q^{2}=-m_{J/\Psi}^2$, where $m_{J/\Psi}$ is the
vector meson mass.

In the computation of the $\eta_{c}$ EFF and TFF in the AM, we use
the {\em ansatz} for the quark-photon vertex as given in Eq.~(8)
of the Ref.~\cite{Raya:2015gva}. It has earlier been employed
successfully in the description of the pion EFF and TFF. Such form
is derived through the  gauge technique. It satisfies the WTI, is
free of kinematic singularities, reduces to the bare vertex in the
free-field limit, and has the same Poincar\'e transformation
properties as the bare vertex.

\section{\label{sec:etac-eff} $\eta_{c}$ Elastic form factor}

Charge-conjugation eigenstates do not have EFF. At the quark-level
of a meson, this can be attributed to the equal and opposite
charge of the quark and the corresponding antiquark. Nonetheless,
by coupling a vector current to the quarks inside, one can measure
a ``form factor'' that gives information about the internal
structure of the state.

The $\eta_{c}$ meson, analogously to the pion, has only one vector
form factor $F_{\eta_{c}}(Q^{2})$, defined by the
$\eta_{c}\gamma^{*}$ vertex
\begin{equation}
\label{eqn:etacff}
\Lambda^{\eta_{c}\gamma^{*}}_{\mu}(P_{i},P_{f};Q)=F_{\eta_{c}}(Q^{2})(P_{f}+P_{i})_{\mu},
\end{equation}
\noindent where $Q=P_{f}-P_{i}$ is the momentum of the virtual
photon, and $F_{\eta_{c}}(Q^{2})$ is the $\eta_{c}$ EFF, the
information carrier of the internal electromagnetic structure of
the bound state. In our approach, the impulse approximation for
the $\eta_{c}\gamma^{*}$ vertex reads
\begin{multline}
\label{eqn:etac-photon-vertex}
\Lambda^{\eta_{c}\gamma^{*}}_{\mu}(P,Q)=2N_{c}\int\df{k}
\Tr\left[\mi\Gamma_{\eta_{c}}(-P_{f})S(k_{2})\right.\\
\left.\mi\Gamma_{\mu}(Q)S(k_{1})\mi\Gamma_{\eta_{c}}(P_{i})S(k)\right],
\end{multline}
\noindent where $P_{i}=P-Q/2$ and $P_{f}=P+Q/2$ are the incoming
and outgoing meson momenta, respectively, $Q=P_{f}-P_{i}$ is the
virtual photon momentum, and the distribution of momentum between
the constituents such that $k_{1}=k+P-Q/2$ and $k_{2}=k+P+Q/2$.
Since the scattering is elastic, $P_{i}^{2}=P_{f}^{2}=-m_{H}^{2}$.
In terms of $P$ and $Q$, these constraints are cast as $P\cdot
Q=0$ and $P^{2}+Q^{2}/4=-m_{H}^{2}$, where $m_{H}$ is the mass of
the bound state.

\begin{figure}[ht]
\includegraphics[width=0.53\textwidth]{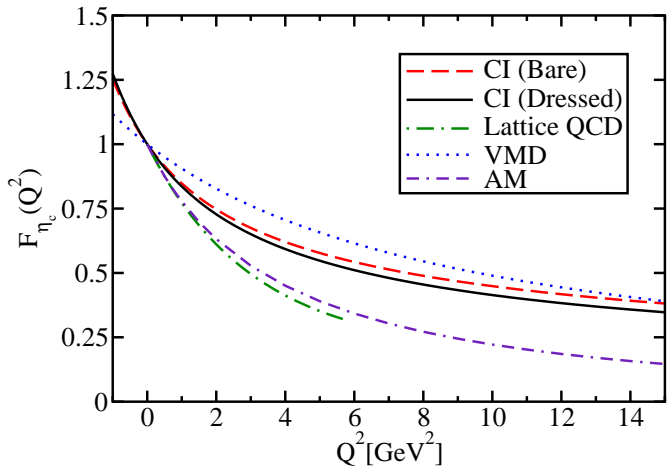}
\caption{\label{fig:etacff} CI results for the $\eta_{c}$ elastic
form factor with and without the dressing of the quark-photon
vertex. The lattice QCD curve is from~\cite{Dudek:2007zz} and the
VMD monopole result is defined by the mass scale $m_{V}=3.096$
GeV. We also include the result obtained with the AM, see
\eqn{eqn:algmod}.}
\end{figure}

In \fig{fig:etacff}, we present our results for the $\eta_{c}$
EFF, displayed with and without the dressing of the quark-photon
vertex. Evidently, this dressing has a negligible effect on the
EFF. In other words, in the CI, the heavy-quark--photon vertex is
almost the bare one for the $Q^{2}$ range shown. However, although
the time-like sector has not been displayed in~\fig{fig:etacff},
the $\eta_{c}$ form factor has a pole at $Q^{2}=-m_{J/\Psi}^{2}$,
where $m_{J/\Psi}$ is the mass of the vector bound state, see
\tab{tab:mcc_all_opt}. This is a consequence of dressing the
quark-photon vertex appropriately. Our results compare well with
the vector meson dominance (VMD) model for $Q^{2}>5\,\GeV^{2}$ but
are harder than the the ones predicted by the lattice QCD. Beyond
$Q^{2} \gtrsim 2.5\,\GeV^{2}$, the CI EFF starts deviating from
the lattice QCD findings by more than $20\%$. Notice that the
lattice QCD curve is a fit to the data~\cite{Dudek:2007zz},
computed in the quenched approximation by assuming the form
$F_{\eta_{c}}(Q^{2})=
\exp\left[-\frac{Q^{2}}{16\beta^{2}}(1+\alpha Q^{2})\right]$,
where $\beta= 0.480(3)$ GeV and $\alpha=-0.046(1)$ GeV$^{-2}$ on
the range $[0,5.5]$ GeV$^{2}$.

\begin{figure}[ht]
\includegraphics[width=0.53\textwidth]{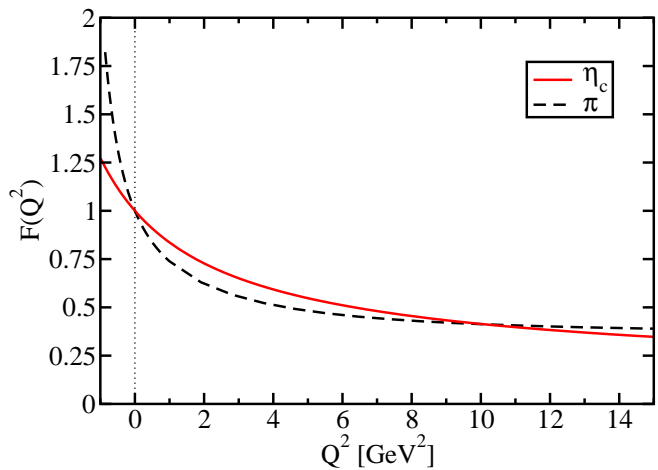}
\caption{\label{fig:etacvspiff}CI results for the $\eta_{c}$ and
$\pi$ elastic form factors. The parameter set used for the
calculation of the $\eta_{c}$ form factor is that used to produce
\tabss{tab:mcc_all_opt}{tab:DecayCNew} and \fig{fig:etacff}, while
the one used to compute the $\pi$ form factor is that given in
~\cite{Roberts:2011wy}.}
\end{figure}

We also compute the $\eta_{c}$ EFF using the AM, defined through
Eqs.~(\ref{Charmonia-am}). Just as for the case of pion EFF, this
model fares very well as compared to the lattice results in all
the range of virtual photon momentum transfer squared $Q^2$. The
large $Q^2$-dependence of the EFF is $1/Q^2$ in contrast with the
corresponding asymptotic behavior obtained through the CI, which
is a constant for $Q^2 \rightarrow \infty$. It is already known
that the asymptotic form factors obtained through the CI are
harder than the QCD
predictions,~\cite{GutierrezGuerrero:2010md,Roberts:2010rn,Roberts:2011wy}.

In \fig{fig:etacvspiff}, we compare the $\eta_{c}$ EFF with that
of the $\pi$, both in the RL approximation with a CI and a dressed
quark-photon vertex. In both cases, the respective EFF tends to a
constant for $Q^{2}\to \infty$, which is a consequence of the
momentum independence of the interaction. Note that the pion EFF
increases more steeply for $Q^{2}<0$ since the pole (the $\rho$
pole for the pion EFF) associated with the dressing of the
light-quark--photon vertex lies very close to $Q^{2}=0$.

The $\eta_{c}$ meson EFF, shown in~\fig{fig:etacvspiff}, can be
parameterized by the following functional form in the $Q^{2}$
range shown
\begin{equation}
\label{eqn:etacffpar}
F_{\eta_{c}}(Q^{2})=\frac{1+ 0.167Q^{2}+ 0.004Q^{4}}{1+ 0.372Q^{2}+ 0.028Q^{4}}.
\end{equation}
\noindent Our values for the $\eta_{c}$ charge radius, defined by
\begin{equation}
\label{eqn:FFradius}
r_{\eta{c}}^{2}=
-6\left.\frac{\mathrm{d}F_{\eta_{c}}(Q^{2})}{\mathrm{d}Q^{2}}\right|_{Q^{2}=0},
\end{equation}
\noindent are presented in \tab{tab:etacchargeradius}. As can be
seen, these compare  well with the ones obtained with the more
sophisticated Maris-Tandy model interaction~\cite{Bhagwat:2006xi},
and with lattice-regularized QCD~\cite{Dudek:2007zz}. Furthermore,
we can compare the charge radius of $\eta_{c}$ with that of $\pi$,
$r_{\pi}=0.45\,\text{fm}$, computed in Ref.~\cite{Roberts:2011wy}.
Obviously, $r_{\eta_{c}}<r_{\pi}$; i.e., the heavier the meson,
the closer it is to being a point particle. We also report the
$\eta_{c}$ charge radius using the algebraic model AM,
Eqs.~(\ref{Charmonia-am}). Expectedly, its prediction lies higher
than the CI model (which produces form factors harder than the
ones computed in real QCD) and is in line with the result of
lattice QCD. Furthermore, the charge radius result, presented in
\tab{tab:etacchargeradius}, is also in excellent agreement with
the lattice QCD.

\begin{table}[h]
\begin{center}
\begin{tabular}{lllll}
\hline \hline
  \multicolumn{5}{c} {$\eta_{c}$ charge radius (fm) } \\
\hline \hline
 SDE~\cite{Bhagwat:2006xi} & Lattice QCD ~\cite{Dudek:2007zz} & VMD & CI &
AM \\
 0.219 & 0.25 & 0.156 & 0.219 (0.210) & 0.256 \\
 \hline \hline
\end{tabular}
\caption{\label{tab:etacchargeradius} The charge radius for the
state $\eta_{c}(1S)$ with the CI and various other calculations.
The parameter set used to produce the CI results is that used for
\tabss{tab:mcc_all_opt}{tab:DecayCNew} and \fig{fig:etacff}. We
also include the VMD model result with a mass scale $m_{V}=3.096$
GeV. The value in parenthesis is obtained when we use the bare
quark-photon vertex. We have also computed the $\eta_{c}$ charge
radius using the AM, Eqs.~(\ref{Charmonia-am}).}
\end{center}
\end{table}

\section{\label{etac-tff} $\gamma\gamma^{*}\to\eta_{c}$ Transition Form
Factor}

The interaction vertex describing the
$\gamma^{*}\gamma\to\eta_{c}$ transition can be parameterized by
just one form factor
$G_{\gamma^{*}\gamma\eta_{c}}(Q_{1}^{2},Q_{2}^{2})$, which can be
computed from

\begin{equation}
\label{etactransitionfull}
\mathcal{T}_{\mu\nu}(Q_{1},Q_{2})=T_{\mu\nu}(Q_{1},Q_{2}) + T_{\nu\mu}(Q_{2},Q_{1}),
\end{equation}
\noindent where $Q_{1}$ and $Q_{2}$ are the incoming photon
momenta, $P=Q_{1}+Q_{2}$ is the $\eta_{c}$ momentum, and
\begin{eqnarray}
\label{etactransition}
T_{\mu\nu}(Q_{1},Q_{2})&=&\frac{\alpha_{\text{em}}}{\pi f_{\eta_{c}}}
\epsilon_{\mu\nu\alpha\beta}Q_{1\alpha}Q_{2\beta}
G_{\gamma^{*}\gamma\eta_{c}}(Q_{1}^{2},Q_{2}^{2}) \nonumber \\
&& \hspace{-1.9cm} \; = \;  \Tr\int\df{k}S(k_{1})\Gamma_{\eta_{c}}(k_{1},k_{2};P)S(k_{2}) \nonumber \\
&& \hspace{-1.9cm} \; \times \;
\mi\mathcal{Q}_{c}\Gamma_{\mu}(k_{2},k_{3};Q_{2})S(k_{3})
\mi\mathcal{Q}_{c}\Gamma_{\nu}(k_{3},k_{1};Q_{1}),
\end{eqnarray}

\noindent with $k_{1}=k-Q_{1}$, $k_{2}=k+Q_{2}$, $k_{3}=k$,
$\mathcal{Q}_{c}=(2/3)e$ and
$\alpha_{\text{em}}={e^{2}}/{(4\pi)}$. The kinematic constraints
are $Q_{1}^{2}=Q^{2}$, $Q_{2}^{2}=0$ and $Q_{1}\cdot
Q_{2}=-(m_{\eta_{c}}^{2}+Q^{2})/2$, where
$P^{2}=-m_{\eta_{c}}^{2}$, with $m_{\eta_{c}}$ the $\eta_{c}$
mass.

\begin{figure}[ht]
\includegraphics[width=0.53\textwidth]{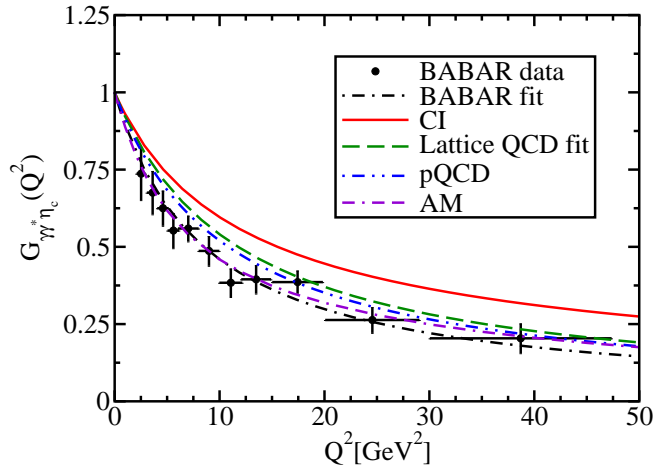}
\caption{\label{fig:etactff}CI for the transition
$\gamma^{*}\gamma\to\eta_{c}$ form factor. The lattice QCD curve
is a fit~\cite{Dudek:2007zz} to data of the form
$G_{\gamma^{*}\gamma\eta_{c}}^{\text{Lattice}}(Q^{2})=\frac{\mu^{2}}{\mu^{2}+Q^{2}}$
with $\mu=3.43$ GeV, while the BABAR data is a
fit~\cite{Druzhinin:2010zza} of the form
$G_{\gamma^{*}\gamma\eta_{c}}^{\text{BABAR}}(Q^{2})=\frac{1}{1+Q^{2}/\Lambda}$
with $\Lambda=8.5$ GeV$^{2}$. The perturbative QCD (pQCD) limit is
due to Feldmann and Kroll~\cite{Feldmann:1997te}. We also include
the plot obtained with the algebraic model, see \eqn{eqn:algmod}.}
\end{figure}

\begin{figure}[ht]
\includegraphics[width=0.53\textwidth]{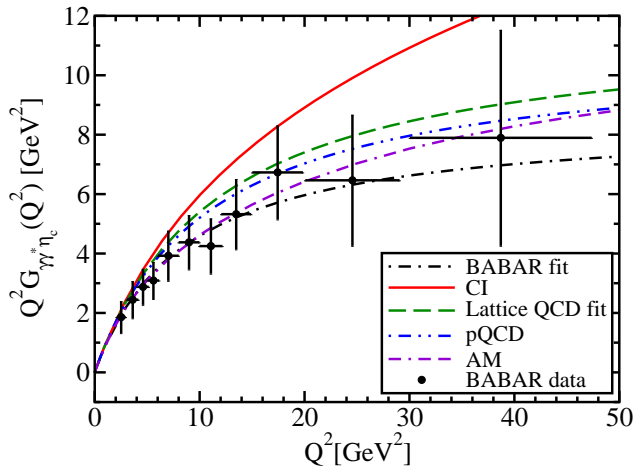}
\caption{\label{fig:etacQ2tff}Numerical results for
$Q^{2}G_{\gamma^{*}\gamma\eta_{c}}$. See caption for
\fig{fig:etactff}.}
\end{figure}

In \fig{fig:etactff}, we present the CI results for the
$\gamma^{*}\gamma\to\eta_{c}$ TFF. Although not shown in
\fig{fig:etactff}, the form factor has a pole at
$Q^{2}=-m_{J/\Psi}^{2}$, where $m_{J/\Psi}$ is the mass of the
vector bound state. The results compare fairly well with the BABAR
data and lattice QCD for low $Q^{2}$. For this reason, the
interaction radius of the transition form factor, defined
in~\eqn{eqn:FFradius} and tabulated
in~\tab{tab:etactchargeradius}, compares well with the lattice QCD
and BABAR findings, as it probes the slope of the TFF for $Q^2 \to
0$. However, for intermediate to large $Q^{2}$, CI provides a
harder form factor and the correct asymptotic $Q^{2}$ behavior is
not captured, see also \fig{fig:etacQ2tff}.

\begin{table}[h]
\begin{center}
\begin{tabular}{llll}
\hline \hline
\multicolumn{4}{c} {interaction radius (fm) } \\
\hline \hline
 BABAR~\cite{Druzhinin:2010zza} & Lattice QCD ~\cite{Dudek:2007zz} & CI &
AM \\
 0.166 & 0.141 & 0.133 & 0.17 \\
 \hline \hline
\end{tabular}
\caption{\label{tab:etactchargeradius} Interaction radius of the
transition $\gamma^{*}\gamma\to\eta_{c}$ form factor as defined in
\eqn{eqn:FFradius}. The BABAR and lattice QCD results were
extracted from their respective monopole parametrization of the
data. We also report the results obtained with the
AM,~\eqn{eqn:algmod}.}
\end{center}
\end{table}

Both the EFF and TFF obtained from the CI and displayed in
\figs{fig:etacff}{fig:etactff} tend to a constant when
$Q^{2}\to\infty$. This is because the quark propagator mass
function and BSAs are momentum independent, which is a
characteristic of the CI, see~\eqn{eqn:contact_interaction}. The
need thus arises to calculate these objects, and the resulting
form factors, with a more realistic interaction, a computation
that is underway and will be reported elsewhere.

 As this is numerically
more demanding, we resort to the AM, defined in
Eq.~(\ref{Charmonia-am}). In \figs{fig:etactff}{fig:etacQ2tff},
the numerical results for $G_{\gamma\gamma^{*}\eta_{c}}$ and
$Q^{2}G_{\gamma\gamma^{*}\eta_{c}}$, respectively, contain the
plots obtained through employing the AM,~\eqn{eqn:algmod},
perturbative QCD calculation~\cite{Feldmann:1997te}, as well as
the BABAR data. As can be seen from these figures, the AM produces
results which agree well with experiment for all the range of
$Q^2$, where results are available. Moreover, it behaves like
$1/Q^{2}$ for large $Q^{2}$ and matches onto the perturbative QCD
limit of the TFF.


\section{\label{sec:PDA} $\eta_c$ Parton Distribution Amplitude}

The perturbative calculation of the $\eta_{c} \rightarrow
\gamma\gamma^{*}$ TFF in~Ref.~\cite{Feldmann:1997te} is based upon
a factorization of short- and long-distance physics. In other
words, it is a convolution of a hard-scattering amplitude computed
perturbatively from QCD and a universal hadronic light-cone wave
function. This wave function cannot be determined completely
accurately but the $\eta_c$ decay constant, which probes the wave
function at origin, can provide stringent constraints on the
latter. On the other hand, the parton distribution amplitude (PDA)
is also connected to the wave function, the former being the
integration of the latter over the transverse momentum. This
interconnection was exploited in~\cite{Feldmann:1997te} to propose
the following parametrization for the $\eta_{c}$ PDA for all
spacelike values of $Q^{2}$:
\begin{figure}[ht]
\includegraphics[width=0.53\textwidth]{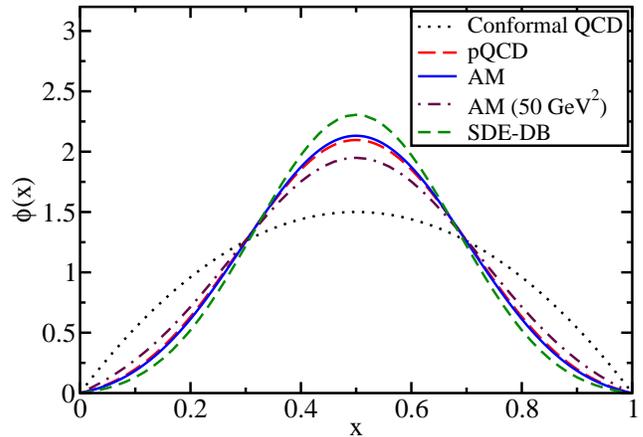}
\caption{\label{fig:etacpda} Numerical results for the $\eta_c$
PDA, $\phi(x)$, obtained with the AM. We have also plotted the
resulting PDA evolved to 50 GeV$^2$, using the leading-order QCD
ERBL evolution equation (see text). For the sake of comparison, we
have also included the perturbative QCD (pQCD) results of
Ref.~\cite{Feldmann:1997te}, the asymptotic QCD expression
$6x(1-x)$ (Conformal QCD), and the recent SDE prediction (SDE-DB)
of Ref.~\cite{Ding:2015rkn}.}
\end{figure}
\begin{equation}
\label{eqn:pda}
\phi(x)=N_{\phi}(a)x(1-x)\exp\left[-a^{2}m_{\eta_{c}}^{2}(x-x_{0})^{2}\right],
\end{equation}
\noindent where $N_{\phi}(a)$ is fixed such that
$\int_{0}^{1}\dx{x}\phi(x)=1$ and $x_{0}=1/2$. The conformal limit
of this meson parton distribution amplitude
$\phi^{asy}(x)=6x(1-x)$ is obtained formally in the limit
$am_{\eta_{c}}\to 0$).

In Ref.~\cite{Feldmann:1997te}, determination of the $\eta_c$ decay
constant suggests the value $a=0.97\,\GeV^{-1}$ for the transverse
size parameter. This value is also consistent with the estimates
for the charge radius squared or the quark velocity in potential
models,~\cite{Buchmuller:1980su}. The behavior of the
PDA,~\eqn{eqn:pda}, plotted in \fig{fig:etacpda}, resembles the
theoretically expected and experimentally confirmed behavior of
heavy hadron fragmentation functions; see~\cite{Feldmann:1997te}
and references therein.

In our current work, we refer to the novel method, developed
in~\cite{Chang:2013pq}, to compute the meson PDA from the
projection onto the light-front of the meson's
Poincar\'e-covariant Bethe-Salpeter wave-function. Carrying out
this exercise for the AM,~\eqn{eqn:algmod}, produces $a\cong
1\,\GeV^{-1}$. Note that the values of $a$ and
$\sigma=m_{\pi}/m_{\eta_c}$, used in defining the AM,
 are
correlated: a small value of $\sigma$ gives $a$ closer to 1 while
a value $\approx 1$ recovers the PDA of the pion obtained with the
AM.

This computation produces plots displayed in \fig{fig:etacpda}. In
the same figure, we also present the AM result for the PDA,
evolved from $4\,\GeV^{2}$ to $50\,\GeV^{2}$ through the
leading-order Efremov-Radyushkin-Brodsky-Lepage (ERBL) evolution
equation,~\cite{Efremov:2009dx,Efremov:1979qk,Lepage:1979zb,
Lepage:1980fj}. Interestingly, the AM result is practically
indistinguishable from the result of~\cite{Feldmann:1997te}. For
the sake of comparison, we also superimpose the result reported
in~\cite{Ding:2015rkn}, obtained with a sophisticated
DCSB-improved SDE truncation (SDE-DB).

Together with the results for the form factors, this analysis
essentially means that the SDE-based AM already encodes a reliable
description of the $\eta_{c}$ meson and that a full numerical
calculation with a realistic interaction should  reproduce similar
results.

\section{\label{sec:conclusions} Conclusions}

We have computed the EFF and the TFF ($\eta_c(1S) \rightarrow
\gamma \gamma^*$) for the $\eta_{c}$ meson, and the corresponding
charge radii, in a CI as well as an SDE-BSE formalism inspired AM.
Within the CI, we employ the dressing of the quark-photon vertex,
consistent with the model truncation and the WTI. It ensures the
form factor possesses a vector meson pole at
$Q^{2}=-m_{J/\Psi}^{2}$. Since the mass $m_{J/\Psi}$ is large, the
effect of the meson vector pole on the charge radii is very small,
i.e., the heavier the meson the closer it is to a point particle.
Our CI is based upon a good description of the masses of the
ground state in four different channels: pseudo-scalar
($\eta_c(1S)$), vector ($J/\Psi(1S)$), scalar ($\chi_{c_0}(1P)$)
and axial vector ($\chi_{c_{1}}(1P)$), as well as the weak decay
constants of the $\eta_c(1S)$ and $J/\Psi(1S)$, and the charge
radius of $\eta_c(1S)$.

For the form factors, expectedly, the CI results agree with QCD
based prediction and/or experiments only up to a certain value of
the virtual photon momentum transfer $Q^2$. This observation is in
line with earlier similar calculations for the $\pi$ and the
$\rho$,\cite{GutierrezGuerrero:2010md,Roberts:2011wy,Roberts:2010rn},
 where it is argued that the form factors of hadrons in a
CI are harder than the real QCD based results. Therefore, for the
CI, both form factors tend to a constant for $Q^{2}\to \infty$,
which is a consequence of the momentum-independent interaction.

Furthermore, we have also extended an SDE-BSE based AM, proposed
for the light quark sector, to study the $\eta_c$. We calculate
EFF, TFF ($\eta_c(1S) \rightarrow \gamma \gamma^*$) and also the
$\eta_{c}$ PDA with this model. For the EFF, the results are in
excellent agreement with the lattice findings for all $Q^2$
available. An extra advantage of the AM is that its simplicity
allows us to extend the computation to any desired values of
space-like $Q^2$. We show the results till $Q^2 = 15$ GeV$^2$ for
the EFF. For the TFF ($\eta_c(1S) \rightarrow \gamma \gamma^*$, we
calculate the results till $Q^2 = 50$ GeV$^2$. For all the regime
of momentum transfer squared $Q^2$, the results match perfectly
with the experiment. Moreover, for large $Q^2$, the perturbative
QCD limit of~\cite{Feldmann:1997te} is faithfully reproduced.

This essentially means that the AM already gives a good
description of the $\eta_{c}$ meson and that a full numerical
calculation with a realistic interaction should be able to produce
similar results. We are currently in the process of extending our
work to the sector of bottomonia. \\

\section{Acknowledgments}

The authors acknowledge financial support from CONACyT (doctoral
scholarship for M.A.~Bedolla and K.~Raya; postdoctoral contract
No.~290917-UMSNH for J.J.~Cobos-Mart\'{\i}nez and research grant
No. CB-2014-242117 for A.~Bashir). This work has also partly been
financed by the CIC-UMSNH grant 4.10.


\bibliography{SDEBSEReferences}

\end{document}